\documentclass[twocolumn,conference]{IEEEtran}
\usepackage[T1]{fontenc}
\usepackage[latin9]{inputenc}
\usepackage[a4paper]{geometry}
\usepackage{xcolor}
\usepackage{pdfcolmk}
\usepackage{amsmath}
\usepackage{amsthm}
\usepackage{amssymb}
\usepackage{stackrel}
\usepackage{graphicx}
\PassOptionsToPackage{normalem}{ulem}
\usepackage{ulem}
\usepackage[unicode=true,
 bookmarks=true,bookmarksnumbered=true,bookmarksopen=true,bookmarksopenlevel=1,
 breaklinks=false,pdfborder={0 0 0},pdfborderstyle={},backref=false,colorlinks=false]
 {hyperref}
\hypersetup{pdftitle={Your Title},
 pdfauthor={Your Name},
 pdfpagelayout=OneColumn, pdfnewwindow=true, pdfstartview=XYZ, plainpages=false}

\makeatletter

\providecolor{lyxadded}{rgb}{0,0,1}
\providecolor{lyxdeleted}{rgb}{1,0,0}

\DeclareRobustCommand{\lyxsout}[1]{\ifx\\#1\else\sout{#1}\fi}

\theoremstyle{plain}
\newtheorem{lem}{\protect\lemmaname}
\theoremstyle{plain}
\newtheorem{cor}{\protect\corollaryname}
\theoremstyle{plain}
\newtheorem{thm}{\protect\theoremname}

\usepackage[caption=false,font=footnotesize]{subfig}

\makeatother

\providecommand{\corollaryname}{Corollary}
\providecommand{\lemmaname}{Lemma}
\providecommand{\theoremname}{Theorem}

\begin{document}
\title{Capacity-CRB Tradeoff in OFDM Integrated Sensing and Communication
Systems}
\author{\IEEEauthorblockN{Zhe Huang\IEEEauthorrefmark{1}, An liu\IEEEauthorrefmark{2}, Rui
Du\IEEEauthorrefmark{3} and Tony Xiao Han\IEEEauthorrefmark{4}}\IEEEauthorblockA{\IEEEauthorrefmark{1}College of Information Science and Electronic
Engineering, Zhejiang University, Zhejiang, China\\
Email: exploer\_huangzhe@zju.edu.cn}\IEEEauthorblockA{\IEEEauthorrefmark{2}College of Information Science and Electronic
Engineering, Zhejiang University, Zhejiang, China\\
Email: anliu@zju.edu.cn}\IEEEauthorblockA{\IEEEauthorrefmark{3}Huawei Technologies Co., Ltd., China\\
Email: ray.du@huawei.com}\IEEEauthorblockA{\IEEEauthorrefmark{4}Huawei Technologies Co., Ltd., China\\
Email: tony.hanxiao@huawei.com}}
\maketitle
\begin{abstract}
Integrated sensing and communication (ISAC) has emerged as a key technology
for future communication systems. In this paper, we provide a general
framework to reveal the fundamental tradeoff between sensing and communication
in OFDM systems, where a unified ISAC waveform is exploited to perform
both tasks. In particular, we define the Capacity-Bayesian Cramer
Rao Bound (BCRB) region in the asymptotically case when the number
of subcarriers is large. Specifically, we show that the asymptotically
optimal input distribution that achieves the Pareto boundary point
of the Capacity-BCRB region is Gaussian and the entire Pareto boundary
can be obtained by solving a convex power allocation problem. Moreover,
we characterize the structure of the sensing-optimal power allocation
in the asymptotically case. Finally, numerical simulations are conducted
to verify the theoretical analysis and provide useful insights.

\end{abstract}

\section{Introduction}

Future 6G communication system will integrate radar sensing and communication
functions to support various important application scenarios, such
as autonomous driving and smart cities \cite{liu2022survey,JRC}.
Under such a background, integrated sensing and communication (ISAC),
in which communication signals are exploited to simultaneously achieve
high-speed communication and high-accuracy sensing, has emerged as
a key technology for future communication systems \cite{liu2022survey,JRC}.

As the first step of the research on ISAC, it is necessary to investigate
the fundamental performance limits of ISAC and to quantify the optimal
tradeoff between sensing accuracy and communication rate for a given
ISAC scenario. Recently, a number of works have been dedicated to
studying the fundamental limits of ISAC, see, e.g., \cite{liu2022information,kobayashi2018joint}.
A notion of capacity-distortion function built on rate-distortion
theory is introduced in \cite{liu2022information,kobayashi2018joint}
for discrete memoryless channels (DMC). The sensing accuracy is quantified
by general distortion functions while the communication performance
is still quantified by classic communication rate. However, the DMC
model is oversimplified to capture the key features of practical ISAC
systems. For example, in practice, the sensing state such as the delay/range
of the target is not i.i.d. and more complicated physical layer technologies
such as MIMO/OFDM are widely used in 5G and beyond systems. Moreover,
it is often difficult to analyze/calculate the exact distortion for
a practical system, and we have to resort to various lower bounds
of the distortion such as the Bayesian Cramer Rao Bound (BCRB) for
tractable analysis. As such, it is practically important to study
the capacity-BCRB tradeoff in MIMO/OFDM ISAC systems \cite{xiong2022flowing,hua2022mimo}.

In this paper, we consider a mono-static OFDM ISAC system, where the
BS serves a mobile user while detecting targets using the same OFDM
waveform. For clarity, we focus on the single-input single-output
(SISO) OFDM ISAC systems since the capacity-BCRB tradeoff for the
SISO case is still open. There are some early attempts to study the
capacity-CRB\textbackslash BCRB tradeoff in MIMO ISAC system, see,
e.g, \cite{xiong2022flowing,hua2022mimo}. However, in \cite{xiong2022flowing}
the authors only capture the optimal input distribution for the communication-optimal
and sensing-optimal boundary points and in \cite{hua2022mimo} the
authors assume the optimal input distribution is Gaussian without
providing a rigorous proof. To overcome these drawbacks, in this paper
we establish the Capacity-BCRB region for OFDM ISAC system in the
asymptotically case when the number of subcarriers is large. Specifically,
we show that the asymptotically optimal input distribution that achieves
the Pareto boundary point of the Capacity-BCRB region is Gaussian
and the entire Pareto boundary can be obtained by solving a convex
power allocation problem. Moreover, we characterized the structure
of the sensing-optimal power allocation in the asymptotically case.
Finally, numerical simulations are conducted to verify the theoretical
analysis and provide useful insights.

\section{System Model and Performance Metrics}

\subsection{System Model}

Consider an OFDM ISAC system with one BS serving a mobile user while
detecting $K$ targets indexed by $k\in\left\{ 1,\ldots,K\right\} $.
The system consists of a single-antenna BS with $N$ subcarriers and
a single-antenna user. Assume a block fading channel model where both
the radar target parameters and communication channel remain constant
within $M$ OFDM symbol durations.

In the $m$-th symbol duration, the BS transmits a frequency domain
symbol $\boldsymbol{x}_{m}\in\mathbb{C}^{N}$, and the corresponding
echo signal in the frequency domain can be expressed as
\begin{equation}
\boldsymbol{y}_{m}^{r}=\textrm{diag}\left(\boldsymbol{x}_{m}\right)\boldsymbol{h}^{r}+\boldsymbol{n}_{m}^{r},\label{eq:downlink_signal}
\end{equation}
where $\boldsymbol{h}^{r}\in\mathbb{C}^{N}$ is the radar channel
vector and $\boldsymbol{n}_{m}^{r}\sim\mathcal{CN}\left(0,\left(\sigma^{r}\right)^{2}\mathbf{I}\right)\in\mathbb{C}^{N}$
is the additive white Gaussian noise (AWGN) with variance $\left(\sigma_{n}^{r}\right)^{2}$.
The radar channel vector depends on the delays and radar cross sections
(RCSs) of the targets and can be modeled as
\begin{equation}
\boldsymbol{h}^{r}=\sum_{k=1}^{K}\alpha_{k}^{r}\boldsymbol{\varphi}\left(\tau_{k}^{r}\right),\label{eq:radarchannel}
\end{equation}
where $\tau_{k}^{r}$ and $\alpha_{k}^{r}$ are the delay and RCS
of the $k$-th target, and $\boldsymbol{\varphi}\left(\tau\right)\in\mathbb{C}^{N}$
is given by

\begin{equation}
\boldsymbol{\varphi}\left(\tau\right)=\left[1,e^{-j2\pi f_{0}\tau},\ldots,e^{-j2\pi\left(N-1\right)f_{0}\tau}\right]^{T},
\end{equation}
where $f_{0}$ is the subcarrier spacing. The received frequency-domain
communication signal at the user can be expressed as
\begin{equation}
\boldsymbol{y}_{m}^{c}=\textrm{diag}\left(\boldsymbol{x}_{m}\right)\boldsymbol{h}^{c}+\boldsymbol{n}_{m}^{c},\label{eq:uplink_signal}
\end{equation}
where $\boldsymbol{h}^{c}\in\mathbb{C}^{N}$ is the communication
channel vector and $\boldsymbol{n}_{m}^{c}\sim\mathcal{CN}\left(0,\left(\sigma^{c}\right)^{2}\mathbf{I}\right)\in\mathbb{C}^{N}$
is the AWGN.

For convenience, define the aggregation of all the unknown sensing
parameters for the $K$ targets as $\boldsymbol{\theta}^{r}\triangleq\left[\left(\boldsymbol{\theta}_{1}^{r}\right)^{T}\cdots\left(\boldsymbol{\theta}_{K}^{r}\right)^{T}\right]^{T}$
and $\boldsymbol{\theta}_{k}^{r}\triangleq\left[\begin{array}{ccc}
\textrm{Re}\left(\alpha_{k}^{r}\right), & \textrm{Im}\left(\alpha_{k}^{r}\right), & \tau_{k}^{r}\end{array}\right]^{T}.$

Define the aggregation of all the transmitted symbols as $\boldsymbol{x}\triangleq\left[\left(\boldsymbol{x}_{1}\right)^{T}\cdots\left(\boldsymbol{x}_{M}\right)^{T}\right]^{T}$,
the aggregation of all the echo signal $\boldsymbol{y}^{r}\triangleq\left[\left(\boldsymbol{y}_{1}^{r}\right)^{T}\cdots\left(\boldsymbol{y}_{M}^{r}\right)^{T}\right]^{T}$
and the aggregation of all the downlink signal $\boldsymbol{y}^{c}\triangleq\left[\left(\boldsymbol{y}_{1}^{c}\right)^{T}\cdots\left(\boldsymbol{y}_{M}^{c}\right)^{T}\right]^{T}$.

We assume that the BS knows the communication channel vector $\boldsymbol{h}^{c}$
from the channel estimation stage and certain prior information $p_{\mathbf{\Theta}^{r}}\left(\boldsymbol{\theta}^{r}\right)$
about the target parameters $\boldsymbol{\theta}^{r}$ from e.g.,
the sensing results in the previous block (assuming the target parameters
of adjacent blocks are correlated according to certain probability
model). The transmitted frequency domain data symbol $\boldsymbol{x}_{m}$
is known to the BS but unknown to the user. Moreover, $\boldsymbol{x}_{m},m=1,...,M$
are \textit{i.i.d.} over different symbols for certain input distribution
$p_{\mathbf{X}}$, i.e. $p_{\mathbf{X}}\left(\boldsymbol{x}\right)=\stackrel[m=1]{M}{\prod}p_{\mathbf{X}}\left(\boldsymbol{x}_{m}\right)$.

Note that for clarity, we focus on the range/delay estimation of the
targets for sensing, and thus it is sufficient to consider a wideband
single-input single-output (SISO) systems since the delay estimation
performance is mainly determined by the system bandwidth. In a wideband
multiple-input-multiple-output (MIMO) ISAC system with Angle of arrival
(AoA) and Doppler estimation capability, the model in (\ref{eq:radarchannel})
should also include the AoA and Doppler of the target's paths, which
is left as future work.

\subsection{Capacity-Distortion and Capacity-CRB Tradeoff}

\begin{figure}[tbh]
\begin{centering}
\includegraphics[width=80mm]{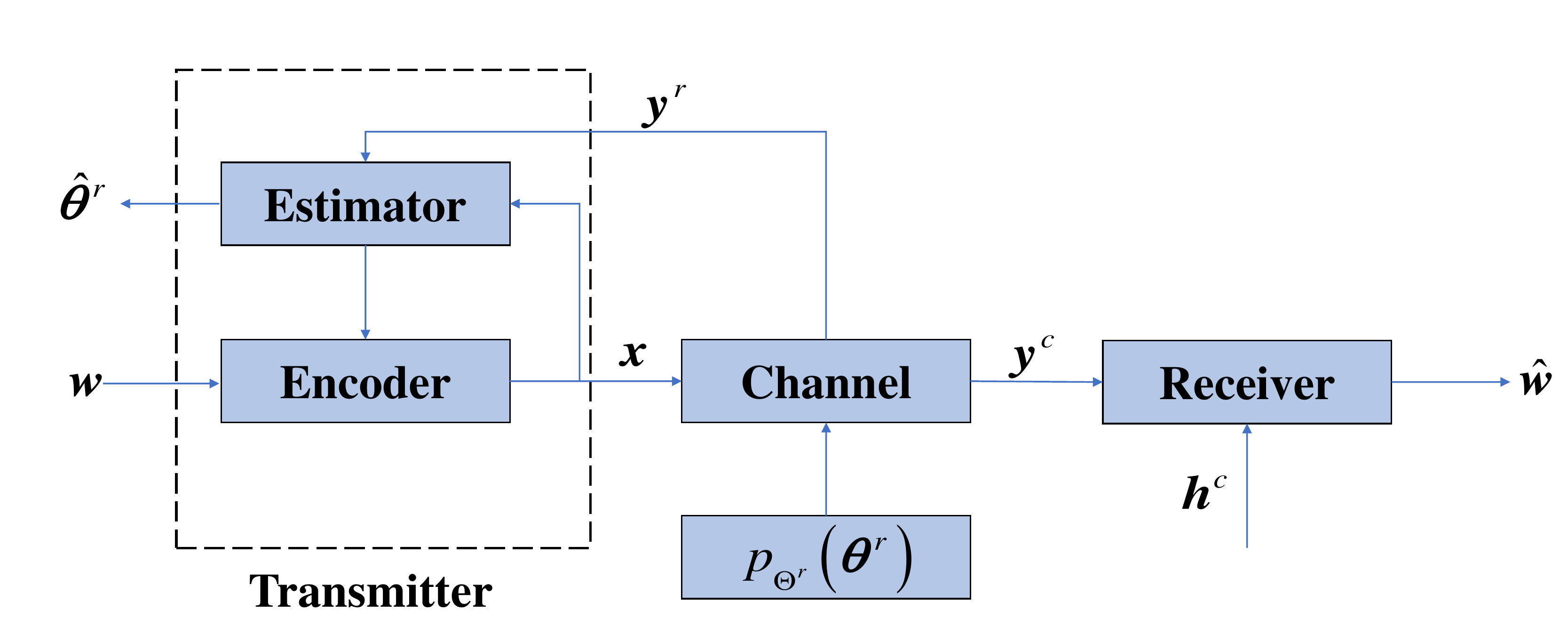}
\par\end{centering}
\caption{\label{fig:channel_model}A point-to-point OFDM-ISAC channel.}
\end{figure}

The capacity-distortion function $C\left(D\right)$ represents the
tradeoff between communication capacity and sensing distortion for
ISAC systems. In this section, we study the optimal capacity-distortion
tradeoff $C\left(D\right)$ for the above OFDM-ISAC system as illustrated
in Fig. \ref{fig:channel_model}. Specifically, we choose the sensing
distortion metric $e\left(\boldsymbol{\theta}^{r},\hat{\boldsymbol{\theta}}^{r}\right)$
as the mean squared error (MSE), i.e., $e\left(\boldsymbol{\theta}^{r},\hat{\boldsymbol{\theta}}^{r}\right)=\left\Vert \boldsymbol{\theta}^{r}-\hat{\boldsymbol{\theta}}^{r}\right\Vert ^{2}$,
where $\hat{\boldsymbol{\theta}}^{r}$ is the estimator of $\boldsymbol{\theta}^{r}$.
In this case, for any given transmit signal $\boldsymbol{x}$ and
received echo signal $\boldsymbol{y}^{r}$, the optimal estimator
for $\boldsymbol{\theta}^{r}$ is given by the minimum MSE (MMSE)
estimator as \cite{kay1993fundamentals}

\begin{equation}
\hat{\boldsymbol{\theta}}^{r}\left(\boldsymbol{y}^{r},\boldsymbol{x}\right)\triangleq\int p\left(\boldsymbol{\theta}^{r}|\boldsymbol{y}^{r},\boldsymbol{x}\right)\boldsymbol{\theta}^{r}d\boldsymbol{\theta}^{r}.
\end{equation}

Following similar analysis as in \cite{liu2022information,kobayashi2018joint},
the optimal capacity-distortion of the OFDM ISAC system considered
is given by

\begin{subequations}
\begin{eqnarray}
\mathcal{P:}C\left(D\right)= & \underset{p_{\mathbf{X}}}{\max} & I\left(\mathbf{Y}^{c};\mathbf{X}\right),\\
 & s.t. & \int p_{\mathbf{X}}\left(\boldsymbol{x}\right)d\boldsymbol{x}=1,\\
 &  & \int p_{\mathbf{X}}\left(\boldsymbol{x}\right)\left\Vert \boldsymbol{x}\right\Vert ^{2}d\boldsymbol{x}\leq P,\\
 &  & \int p_{\mathbf{X}}\left(\boldsymbol{x}\right)c\left(\boldsymbol{x}\right)d\boldsymbol{x}\leq D,\label{eq:distortion_con}
\end{eqnarray}
\end{subequations}

where $I\left(\mathbf{Y}^{c};\mathbf{X}\right)$ is the mutual information
between $\mathbf{Y}^{c}$ and $\mathbf{X}$, $P$ is the total transmitted
power, $D$ is the maximum tolerated distortion, $c\left(\boldsymbol{x}\right)\triangleq\iint p\left(\boldsymbol{y}^{r}|\boldsymbol{x},\boldsymbol{\theta}^{r}\right)p_{\mathbf{\Theta}^{r}}\left(\boldsymbol{\theta}^{r}\right)e\left(\boldsymbol{\theta}^{r},\hat{\boldsymbol{\theta}}^{r}\right)d\boldsymbol{y}^{r}d\boldsymbol{\theta}^{r}$
is the average MSE for given transmit signal $\boldsymbol{x}$, and
the joint distribution of random variables $\left\{ \mathbf{Y}^{r}\mathbf{Y}^{c}\mathbf{X}\mathbf{\Theta}^{r}\right\} $
is given by
\begin{equation}
p_{\mathbf{\Theta}^{r}}\left(\boldsymbol{\theta}^{r}\right)p_{\mathbf{X}}\left(\boldsymbol{x}\right)p\left(\boldsymbol{y}^{c}\mid\boldsymbol{x}\right)p\left(\boldsymbol{y}^{r}\mid\boldsymbol{x},\boldsymbol{\theta}^{r}\right),
\end{equation}
where $p\left(\boldsymbol{y}^{r}\mid\boldsymbol{x},\boldsymbol{\theta}^{r}\right)$
and $p\left(\boldsymbol{y}^{c}\mid\boldsymbol{x}\right)$ are the
Gaussian channel transition probabilities determined by (\ref{eq:downlink_signal})
and (\ref{eq:uplink_signal}), respectively.

In general, it is difficult to obtain the closed-form expressions
of the MMSE estimator and the relevant MMSE. It is well-known that
certain practical estimators are capable of approaching the Bayesian
Cramer-Rao Bound (BCRB) when the $\textrm{SNR}^{r}\triangleq P\left(\sigma^{r}\right)^{-2}$
is sufficiently high \cite{kay1993fundamentals}, so we choose the
BCRB as an alternative way to evaluate the distortion. Therefore,
we can replace the term $c\left(\boldsymbol{x}\right)$ with its BCRB
$\tilde{c}\left(\boldsymbol{x}\right)$ as 
\begin{equation}
\tilde{c}\left(\boldsymbol{x}\right)\triangleq\textrm{Tr}\left\{ \left[\mathbb{E}_{\mathbf{\Theta}^{r}}\left[\mathbf{J}^{o}\left(\boldsymbol{x}\right)\right]+\mathbf{J}^{p}\right]^{-1}\right\} ,\label{eq:BCRB}
\end{equation}
where the Fisher Information Matrices (FIMs) $\mathbf{J}^{o}\left(\boldsymbol{x}\right)$
and $\mathbf{J}^{p}$ are given by

\begin{equation}
\mathbf{J}^{o}\left(\boldsymbol{x}\right)\triangleq\mathbb{E}_{\mathbf{N}^{r}}\left[\boldsymbol{g}^{o}\left(\boldsymbol{\theta}^{r}\right)\boldsymbol{g}^{o}\left(\boldsymbol{\theta}^{r}\right)^{T}\right],
\end{equation}

\begin{equation}
\mathbf{J}^{p}\triangleq\mathbb{E}_{\mathbf{\Theta}^{r}}\left[\boldsymbol{g}^{p}\left(\boldsymbol{\theta}^{r}\right)\boldsymbol{g}^{p}\left(\boldsymbol{\theta}^{r}\right)^{T}\right],\label{eq:pri_J}
\end{equation}
where $\boldsymbol{g}^{o}\left(\boldsymbol{\theta}^{r}\right)\triangleq\nabla_{\boldsymbol{\theta}^{r}}\left(\ln p\left(\boldsymbol{y}^{r}\mid\boldsymbol{\theta}^{r},\boldsymbol{x}\right)\right),\boldsymbol{g}^{p}\left(\boldsymbol{\theta}^{r}\right)\triangleq\nabla_{\boldsymbol{\theta}^{r}}\left(\ln p_{\mathbf{\Theta}^{r}}\left(\boldsymbol{\theta}^{r}\right)\right)$.
Then the capacity-BCRB tradeoff $\tilde{C}\left(D\right)$ is given
by replacing the $c\left(\boldsymbol{x}\right)$ in (\ref{eq:distortion_con})
with its BCRB $\tilde{c}\left(\boldsymbol{x}\right)$.

Note that the exact CRB in (\ref{eq:BCRB}) exhibits a rather complicated
dependence on the frequency domain symbol $\boldsymbol{x}$. Thus,
it does not provide immediate insights on the estimation accuracy.
To remedy this problem, we derive the asymptotic CRB (ACRB) as $N\rightarrow\infty$
in the next section. The latter will be a much simpler function of
the frequency domain symbol $\boldsymbol{x}$. Therefore, the ACRB
is an interesting tool to evaluate the influence of the frequency
domain symbol $\boldsymbol{x}$ on the estimation performance.

\section{Asymptotic Cramer Rao Bound Analysis}

In this section, we derive the exact BCRB and the ACRB.

\subsection{Derivation of the exact BCRB}

By definition, the $\mathbf{J}^{o}\left(\boldsymbol{x}\right)$ can
be decomposed as

\begin{equation}
\mathbf{J}^{o}\left(\boldsymbol{x}\right)=\left[\begin{array}{ccc}
\mathbf{J}{}_{11}^{o}\left(\boldsymbol{x}\right) & \cdots & \mathbf{J}_{1K}^{o}\left(\boldsymbol{x}\right)\\
\vdots & \ddots & \vdots\\
\mathbf{J}_{1K}^{o}\left(\boldsymbol{x}\right)^{T} & \cdots & \mathbf{J}_{KK}^{o}\left(\boldsymbol{x}\right)
\end{array}\right],
\end{equation}
where $\mathbf{J}_{kl}^{o}\left(\boldsymbol{x}\right)\triangleq\underset{m}{\sum}\left(\mathbf{J}_{mkl}^{o}\left(\boldsymbol{x}\right)\right),\forall k,l$
and the $\left(p,q\right)$-th element of the submatrix $\mathbf{J}_{mkl}^{o}\left(\boldsymbol{x}\right)$
is given by

\begin{equation}
\left[\mathbf{J}_{mkl}^{o}\left(\boldsymbol{x}\right)\right]_{pq}=2\left(\sigma^{r}\right)^{-2}\textrm{Re}\left[\left(\frac{\partial\boldsymbol{\mu}_{m}^{r}}{\partial\theta_{kp}^{r}}\right)^{H}\left(\frac{\partial\boldsymbol{\mu}_{m}^{r}}{\partial\theta_{lq}^{r}}\right)\right],
\end{equation}
where $\boldsymbol{\mu}_{m}^{r}\triangleq\textrm{diag}\left(\boldsymbol{x}_{m}\right)\boldsymbol{h}^{r}$
and $p,q\in\left\{ 1,2,3\right\} $. After some straightforward calculation,
the $\left(k,k\right)$-th submatrix $\mathbf{J}_{mkk}^{o}\left(\boldsymbol{x}\right)$
and $\left(k,l\right)$-th submatrix $\mathbf{J}_{mkl}^{o}\left(\boldsymbol{x}\right)$
are given in by (\ref{eq:J_k_k}) and (\ref{eq:J_k_l}).

\begin{equation}
\mathbf{J}_{mkk}^{o}\left(\boldsymbol{x}\right)=2\left(\sigma^{r}\right)^{-2}\left[\begin{array}{cc}
\left(\mathbf{J}_{mkk}^{o}\right)_{11} & \left(\mathbf{J}_{mkk}^{o}\right)_{12}\\
\left(\mathbf{J}_{mkk}^{o}\right)_{12}^{T} & \left(\mathbf{J}_{mkk}^{o}\right)_{22}
\end{array}\right],\label{eq:J_k_k}
\end{equation}

\begin{equation}
\mathbf{J}_{mkl}^{o}\left(\boldsymbol{x}\right)=2\left(\sigma^{r}\right)^{-2}\left[\begin{array}{cc}
\left(\mathbf{J}_{mkl}^{o}\right)_{11} & \left(\mathbf{J}_{mkl}^{o}\right)_{12}\\
\left(\mathbf{J}_{mkl}^{o}\right)_{21} & \left(\mathbf{J}_{mkl}^{o}\right)_{22}
\end{array}\right],\label{eq:J_k_l}
\end{equation}
The sub-matrices in $\mathbf{J}_{mkk}^{o}\left(\boldsymbol{x}\right)$
and $\mathbf{J}_{mkl}^{o}\left(\boldsymbol{x}\right)$ are given as

\begin{subequations}
\begin{eqnarray}
\left(\mathbf{J}_{mkk}^{o}\right)_{11} & \triangleq & \textrm{diag}\left(U_{mk},U_{mk}\right),\\
\left(\mathbf{J}_{mkk}^{o}\right)_{12} & \triangleq & \left[\textrm{Im}\left(\alpha_{k}^{r}T_{mk}\right),-\textrm{Re}\left(\alpha_{k}^{r}T_{mk}\right)\right]^{T},\\
\left(\mathbf{J}_{mkk}^{o}\right)_{22} & \triangleq & \omega_{0}^{2}\left|\alpha_{k}^{r}\right|^{2}V_{mk},\\
\left(\mathbf{J}_{mkl}^{o}\right)_{12} & \triangleq & \left[\textrm{Im}\left(\alpha_{l}^{r}T_{mkl}\right),-\textrm{Re}\left(\alpha_{l}^{r}T_{mkl}\right)\right]^{T},\\
\left(\mathbf{J}_{mkl}^{o}\right)_{21} & \triangleq & \left[\textrm{Im}\left(\alpha_{k}^{r}T_{mkl}\right),-\textrm{Re}\left(\alpha_{k}^{r}T_{mkl}\right)\right]^{T},\\
\left(\mathbf{J}_{mkl}^{o}\right)_{22} & \triangleq & \omega_{0}^{2}\left(\alpha_{k}^{r}\right)^{*}\alpha_{l}^{r}V_{mkl},\\
\left(\mathbf{J}_{mkl}^{o}\right)_{11} & \triangleq & \left[\begin{array}{cc}
\textrm{Re}\left(U_{mkl}\right) & -\textrm{Im\ensuremath{\left(U_{mkl}\right)}}\\
\textrm{Im}\left(U_{mkl}\right) & \textrm{Re}\left(U_{mkl}\right)
\end{array}\right],
\end{eqnarray}
\end{subequations}

where $\omega_{0}\triangleq2\pi f_{0}$.

The collection of $\left\{ U_{mk},U_{mkl},T_{mk},T_{mkl},V_{mk},V_{mkl}\right\} _{m,k,l}$
are given as

\begin{subequations}
\begin{eqnarray}
U_{mk} & = & \underset{n}{\sum}\left|x_{nm}\right|^{2},\\
U_{mkl} & = & \underset{n}{\sum}\left|x_{nm}\right|^{2}e^{j\omega_{0}\left(n-1\right)\bar{\tau}_{kl}^{r}},\\
T_{mk} & = & \underset{n}{\sum}\left(n-1\right)\left|x_{nm}\right|^{2},\\
T_{mkl} & = & \underset{n}{\sum}\left(n-1\right)\left|x_{nm}\right|^{2}e^{j\omega_{0}\left(n-1\right)\bar{\tau}_{kl}^{r}}\\
V_{mk} & = & \underset{n}{\sum}\left(n-1\right)^{2}\left|x_{nm}\right|^{2},\\
V_{mkl} & = & \underset{n}{\sum}\left(n-1\right)^{2}\left|x_{nm}\right|^{2}e^{j\omega_{0}\left(n-1\right)\bar{\tau}_{kl}^{r}},
\end{eqnarray}
\end{subequations}

where $\bar{\tau}_{kl}^{r}\triangleq\tau_{k}^{r}-\tau_{l}^{r}$.

\subsection{Derivation of the ACRB}

It can be observed that the elements in $\mathbf{J}_{mkk}^{o}\left(\boldsymbol{x}\right)$
and $\mathbf{J}_{mkl}^{o}\left(\boldsymbol{x}\right)$ relies on the
realizations of the random variable $\boldsymbol{x}_{m}$ which is
hard to evaluate. Motivated by the classical strong law of large numbers
(SLLN) that the sample mean of \textit{i.i.d }random variables converges
to its mean almost surely, we seek to derive the asymptotic behaviors.
Note that the \textit{i.i.d} assumption may not be valid since the
sample from different subcarriers may be correlated. To overcome this
technical challenge, let us introduce the following lemma, which is
proved in \cite{petrov2009stability}.
\begin{lem}
\label{lem:Strong_LN}Let $\left\{ w_{n}\right\} $ denotes a non-negative
deterministic sequence and $\left\{ Z_{n}\right\} $ denotes a random
non-negative sequence with finite variance. Define

\[
W_{N}\triangleq\underset{n}{\sum}w_{n},T_{N}\triangleq\underset{n}{\sum}w_{n}Z_{n}.
\]
Under the technical conditions claimed in \cite{petrov2009stability},
we have the following generalized SLLN holds

\begin{equation}
\frac{T_{N}-\mathbb{E}\left[T_{N}\right]}{W_{N}}\rightarrow0,a.s.
\end{equation}
\end{lem}
In order to obtain the following corollary, we shall also have some
assumptions on $P_{nm}$ which is given as follow,

\begin{equation}
0\leq P_{nm}\leq P_{max},\forall n,m\label{eq:P_nm_order}
\end{equation}
where $P_{nm}=\mathbb{E}\left[\left|x_{nm}\right|^{2}\right]$ denotes
the transmitted power on $n$-th subcarrier in the $m$-th symbol
duration. Note that the constraint $P_{nm}\leq P_{max}$ for certain
$P_{max}$ with order $O\left(\frac{P}{NM}\right)$ is widely used
in practice, because such a constraint will ensure the non-zero communication
rate at each subcarrier for communication and avoid impractical power
allocation by directly minimizing the CRB without any constraint for
sensing \cite{bicua2018radar,liyanaarachchi2021optimized}.

Substituting $Z_{n}=\left|x_{nm}\right|^{2}$, $w_{n}^{s}\triangleq\left(n-1\right)^{s},s\in\left\{ 0,1,2\right\} $
into the generalized SLLN in Lemma \ref{lem:Strong_LN}, we can prove
the following Corollary. The detailed proof is omitted due to space
limit.
\begin{cor}
Let $w_{n}^{s}\triangleq\left(n-1\right)^{s}$, where $s\in\left\{ 0,1,2\right\} $.
We have

\begin{subequations}
\begin{eqnarray}
 & \frac{U_{mk}-\bar{U}_{mk}}{\underset{n}{\sum}w_{n}^{0}}\rightarrow0, & a.s.\label{eq:U_as}\\
 & \frac{T_{mk}-\bar{T}_{mk}}{\underset{n}{\sum}w_{n}^{1}}\rightarrow0, & a.s.\label{eq:T_as}\\
 & \frac{V_{mk}-\bar{V}_{mk}}{\underset{n}{\sum}w_{n}^{2}}\rightarrow0, & a.s.\label{eq:V_as}
\end{eqnarray}
\end{subequations}

where $\bar{U}_{mk}\triangleq\mathbb{E}\left[U_{mk}\right]=\underset{n}{\sum}w_{n}^{0}P_{nm}$,
$\bar{T}_{mk}\triangleq\mathbb{E}\left[T_{mk}\right]=\underset{n}{\sum}w_{n}^{1}P_{nm}$,
$\bar{V}_{mk}\triangleq\mathbb{E}\left[V_{mk}\right]=\underset{n}{\sum}w_{n}^{2}P_{nm}$
.
\end{cor}
Similarly, let $\tilde{w}_{n}^{s}\triangleq\left(n-1\right)^{s}e^{j\omega_{0}\left(n-1\right)\bar{\tau}_{kl}^{r}}$.
We have

\begin{subequations}
\begin{eqnarray}
 & \frac{U_{mkl}-\bar{U}_{mkl}}{\underset{n}{\sum}\tilde{w}_{n}^{0}}\rightarrow0, & a.s.\\
 & \frac{T_{mkl}-\bar{T}_{mkl}}{\underset{n}{\sum}\tilde{w}_{n}^{1}}\rightarrow0, & a.s.\\
 & \frac{V_{mkl}-\bar{V}_{mkl}}{\underset{n}{\sum}\tilde{w}_{n}^{2}}\rightarrow0, & a.s.
\end{eqnarray}
\end{subequations}

where $\bar{U}_{mkl}\triangleq\mathbb{E}\left[U_{mkl}\right]=\underset{n}{\sum}\tilde{w}_{n}^{0}P_{nm}$,
$\bar{T}_{mkl}\triangleq\mathbb{E}\left[T_{mkl}\right]=\underset{n}{\sum}\tilde{w}_{n}^{1}P_{nm}$,
$\bar{V}_{mkl}\triangleq\mathbb{E}\left[V_{mkl}\right]=\underset{n}{\sum}\tilde{w}_{n}^{2}P_{nm}$.

Define the asymptotic normalized FIM (ANFIM) as $\mathbf{\bar{J}}^{o}\left(\boldsymbol{x}\right)\triangleq\underset{N\rightarrow\infty}{\lim}\mathbf{L}\mathbf{J}^{o}\left(\boldsymbol{x}\right)\mathbf{L}^{T}$,
where
\begin{equation}
\mathbf{L}\triangleq\textrm{blkdiag}\left(\mathbf{L}_{1},\cdots,\mathbf{L}_{K}\right)\text{,}
\end{equation}
with $\mathbf{L}_{k}\triangleq\textrm{diag}\left(N^{-\frac{1}{2}}M^{-\frac{1}{2}},N^{-\frac{1}{2}}M^{-\frac{1}{2}},N^{-\frac{3}{2}}M^{-\frac{1}{2}}\right),\forall k$.
\begin{cor}
Under assumptions in (\ref{eq:P_nm_order}), the ANFIM is given by
(\ref{eq:J_lim}), where $\mathbf{\bar{J}}_{kl}^{o}\left(\boldsymbol{x}\right)\triangleq\underset{m}{\sum}\left(\mathbf{\bar{J}}_{mkl}^{o}\left(\boldsymbol{x}\right)\right),\forall m,k,l$
and $\mathbf{\bar{J}}_{mkl}^{o}\left(\boldsymbol{x}\right)\triangleq\underset{N\rightarrow\infty}{\lim}\mathbf{L}_{k}\mathbf{J}_{mkl}^{o}\left(\boldsymbol{x}\right)\mathbf{L}_{l}^{T},\forall m,k,l$.
Moreover, the Big O order of elements in $\mathbf{\bar{J}}_{kk}^{o}\left(\boldsymbol{x}\right)$
and $\mathbf{\bar{J}}_{kl}^{o}\left(\boldsymbol{x}\right),\forall k,l,p,q$
are given by equation (\ref{eq:asCRB_obv_k-1}) -(\ref{eq:asCRB_obv_k-5})
and (\ref{eq:asCRB_obv_k_l}).
\end{cor}
\begin{equation}
\mathbf{\bar{J}}^{o}\left(\boldsymbol{x}\right)\triangleq\left[\begin{array}{ccc}
\mathbf{\bar{J}}_{11}^{o}\left(\boldsymbol{x}\right) & \cdots & \mathbf{\bar{J}}_{1K}^{o}\left(\boldsymbol{x}\right)\\
\vdots & \ddots & \vdots\\
\mathbf{\bar{J}}_{1K}^{o}\left(\boldsymbol{x}\right)^{T} & \cdots & \mathbf{\bar{J}}_{KK}^{o}\left(\boldsymbol{x}\right)
\end{array}\right]\text{,}\label{eq:J_lim}
\end{equation}

\begin{subequations}
\begin{eqnarray}
O\left[\mathbf{\bar{J}}_{kk}^{o}\left(\boldsymbol{x}\right)\right]_{11} & = & O\left[\mathbf{\bar{J}}_{kk}^{o}\left(\boldsymbol{x}\right)\right]_{22}=P_{max},\label{eq:asCRB_obv_k-1}\\
O\left[\mathbf{\bar{J}}_{kk}^{o}\left(\boldsymbol{x}\right)\right]_{12} & = & O\left[\mathbf{\bar{J}}_{kk}^{o}\left(\boldsymbol{x}\right)\right]_{21}=0,\\
O\left[\mathbf{\bar{J}}_{kk}^{o}\left(\boldsymbol{x}\right)\right]_{13} & = & O\left[\mathbf{\bar{J}}_{kk}^{o}\left(\boldsymbol{x}\right)\right]_{13}=\frac{1}{2}\tilde{\alpha}_{k}^{r}P_{max},\\
O\left[\mathbf{\bar{J}}_{kk}^{o}\left(\boldsymbol{x}\right)\right]_{23} & = & O\left[\mathbf{\bar{J}}_{kk}^{o}\left(\boldsymbol{x}\right)\right]_{32}=-\frac{1}{2}\bar{\alpha}_{k}^{r}P_{max},\label{eq:}\\
O\left[\mathbf{\bar{J}}_{kk}^{o}\left(\boldsymbol{x}\right)\right]_{33} & = & \frac{1}{3}\left(\left|\tilde{\alpha}_{k}^{r}\right|^{2}+\left|\bar{\alpha}_{k}^{r}\right|^{2}\right)P_{max},\label{eq:asCRB_obv_k-5}
\end{eqnarray}
\end{subequations}

\begin{equation}
O\left\{ \frac{\left[\mathbf{\bar{J}}_{kl}^{o}\left(\boldsymbol{x}\right)\right]_{pq}}{\left[\mathbf{\bar{J}}_{kk}^{o}\left(\boldsymbol{x}\right)\right]_{pq}}\right\} =\frac{1}{N},\forall p,q\in\left\{ 1,2,3\right\} ,\label{eq:asCRB_obv_k_l}
\end{equation}
where $\tilde{\alpha}_{k}^{r}\triangleq\omega_{0}\textrm{Im}\left(\alpha_{k}^{r}\right)$
and $\bar{\alpha}_{k}^{r}\triangleq\omega_{0}\textrm{Re}\left(\alpha_{k}^{r}\right)$.
\begin{IEEEproof}
Equation (\ref{eq:asCRB_obv_k-1}) - (\ref{eq:asCRB_obv_k_l}) are
derived using the fact that for any $t\in\left\{ 0,1,2\right\} $,
the Big O order for the general expression of elements in $\mathbf{\bar{J}}_{kk}^{o}\left(\boldsymbol{x}\right)$
and $\mathbf{\bar{J}}_{kl}^{o}\left(\boldsymbol{x}\right),\forall k,l,p,q$
is given by (\ref{eq:order_sum}) \cite{besson2003parameter}, where
$\omega_{kl}\triangleq\omega_{0}\left(\tau_{k}^{r}-\tau_{l}^{r}\right)$.
The other calculation is straightforward and omitted due to space
limit.

\begin{figure*}[tbh]
\begin{equation}
O\left\{ \frac{1}{N^{t}MP_{max}}\underset{n}{\sum}\underset{m}{\sum}\left(n-1\right)^{t}P_{nm}e^{j\omega_{kl}\left(n-1\right)}\right\} =\begin{cases}
\frac{1}{t+1}+O\left(\frac{1}{N}\right), & \omega_{kl}=0,k=l\\
O\left(\frac{1}{N}\right), & \omega_{kl}\neq0,k\neq l
\end{cases},\label{eq:order_sum}
\end{equation}
\end{figure*}
\end{IEEEproof}
Therefore, $\mathbf{\bar{J}}^{o}\left(\boldsymbol{x}\right)$ will
tend to be a block-diagonal matrix with negligible approximation error
as shown in (\ref{eq:error_J}).

\begin{equation}
\underset{N\rightarrow\infty}{\lim}\frac{\left\Vert \mathbf{\bar{J}}^{o}\left(\boldsymbol{x}\right)-\tilde{\mathbf{J}}^{o}\left(\boldsymbol{x}\right)\right\Vert _{F}^{2}}{\left\Vert \mathbf{\bar{J}}^{o}\left(\boldsymbol{x}\right)\right\Vert _{F}^{2}}=0,\label{eq:error_J}
\end{equation}
where $\tilde{\mathbf{J}}^{o}\left(\boldsymbol{x}\right)\triangleq\textrm{blkdiag}\left[\mathbf{\bar{J}}_{11}^{o}\left(\boldsymbol{x}\right),\cdots,\mathbf{\bar{J}}_{KK}^{o}\left(\boldsymbol{x}\right)\right]$.

It can be observed that $\tilde{\mathbf{J}}^{o}\left(\boldsymbol{x}\right)$
does not depend on exact input distribution $p_{\mathbf{X}}\left(\boldsymbol{x}\right)$
but only depends on the power allocation of all subcarriers. Therefore,
we can rewrite $\tilde{\mathbf{J}}^{o}\left(\boldsymbol{x}\right)$
as a function of the aggregated power allocation vector $\boldsymbol{p}\triangleq\left[\boldsymbol{p}_{1}^{T},\ldots,\boldsymbol{p}_{M}^{T}\right]^{T}$:
$\tilde{\mathbf{J}}^{o}\left(\boldsymbol{p}\right)$, where $\boldsymbol{p}_{m}\triangleq\left[P_{1m},\ldots,P_{nm}\right]^{T}$is
the power allocation vector for the $m$-th symbol. Then we are ready
to define the ACRB as a function of $\boldsymbol{p}$ :
\begin{equation}
\textrm{ACRB}\left(\boldsymbol{p}\right)\triangleq\textrm{Tr}\left\{ \left[\mathbb{E}_{\mathbf{\Theta}^{r}}\left[\hat{\mathbf{J}}\left(\boldsymbol{x}\right)\right]+\mathbf{J}^{p}\left(\boldsymbol{\theta}^{r}\right)\right]^{-1}\right\} .\label{eq:aCRB}
\end{equation}
where $\hat{\mathbf{J}}\left(\boldsymbol{x}\right)\triangleq\mathbf{L}^{-T}\tilde{\mathbf{J}}^{o}\left(\boldsymbol{x}\right)\mathbf{L}^{-1}$.
From the above analysis, it is easy to see that ACRB is a good approximation
of BCRB for large $N$. Therefore, for large $N$, we may study the
capacity-ACRB tradeoff as a good approximation for the capacity-BCRB
tradeoff function.

\section{Capacity-BCRB Region Analysis}

\begin{figure}[tbh]
\begin{centering}
\includegraphics[width=80mm]{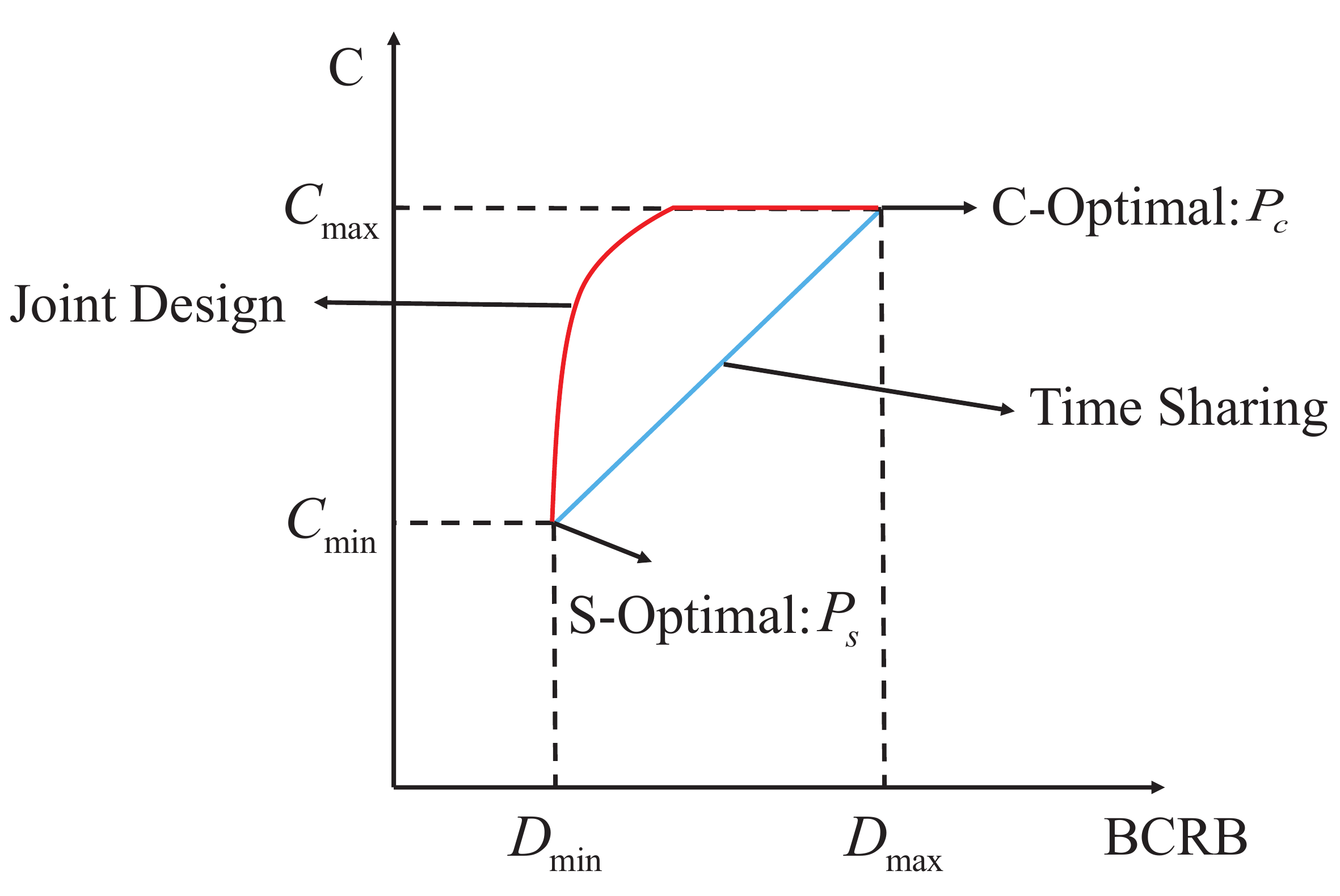}
\par\end{centering}
\caption{\label{fig:C_D_Region}The Capacity-BCRB Regions for OFDM-ISAC system}
\end{figure}

In general, the capacity-BCRB region takes the form illustrated in
Fig. \ref{fig:C_D_Region}. The convex hull of the points $\left(+\infty,0\right),\left(+\infty,C_{max}\right),\left(D_{min},0\right)$
constitutes an inner bound of the capacity-distortion region, where
$P_{s}\triangleq\left(D_{min},C_{min}\right)$ is the sensing-optimal
(S-optimal) point and $P_{c}\triangleq\left(D_{max},C_{max}\right)$
is the communication-optimal (C-optimal) point. The line segment connecting
$P_{s}$ and $P_{c}$ can be achieved using the time-sharing scheme
which allocates orthogonal resources for sensing and communication.

In general, it is very hard to capture the exact boundary of the capacity-BCRB
region (i.e., the red boundary in Fig. \ref{fig:C_D_Region}) for
finite $N$. In the following theorem, we characterize the asymptotically
optimal input distribution that can achieve the boundary of the capacity-BCRB
region for sufficient large $N$ ($N\rightarrow\infty$).
\begin{thm}
\label{thm:asy-optimal-distribution}As $N\rightarrow\infty$, the
asymptotic optimal input distribution $p_{\mathbf{X}}\left(\boldsymbol{x}\right)$
for the optimization problem $\mathcal{P}$ is Gaussian distribution
with zero mean, and thus Problem $\mathcal{P}$ reduces to a power
allocation problem as

\begin{subequations}
\begin{eqnarray}
\bar{\mathcal{P}}:\bar{C}\left(D\right)= & \underset{\boldsymbol{p}}{\max} & f\left(\boldsymbol{p}\right),\\
 & \text{s.t.} & l\left(\boldsymbol{p}\right)\triangleq\mathbf{1}^{T}\boldsymbol{p}\leq P,\\
 &  & d\left(\boldsymbol{p}\right)\triangleq\textrm{\textrm{ACRB}}\left(\boldsymbol{p}\right)\leq D,\\
 &  & 0\leq P_{nm}\leq P_{max},
\end{eqnarray}
\end{subequations}

where $\boldsymbol{p}$ is the aggregated power allocation vector,
and $f\left(\boldsymbol{p}\right)\triangleq\underset{m}{\sum}\underset{n}{\sum}\log_{2}\left(1+\left(\sigma^{c}\right)^{-2}P_{nm}\left|h_{nm}^{c}\right|^{2}\right)$
is the mutual information $I\left(\mathbf{Y}^{c};\mathbf{X}\right)$
under the Gaussian input distribution. Moreover, the power allocation
problem $\bar{\mathcal{P}}$ is convex.
\end{thm}
\begin{IEEEproof}
It can be observed from (\ref{eq:aCRB}) that as $N\rightarrow\infty$,
the sensing performance only depends on the power allocation of all
subcarriers but does not depend the exact input distribution $p_{\mathbf{X}}\left(\boldsymbol{x}\right)$.
On the other hand, for given power allocation, Gaussian input distribution
is optimal for communications. Therefore, the asymptotic optimal input
distribution for the optimization problem $\mathcal{P}$ is Gaussian
and thus $\mathcal{P}$ reduces to the power allocation problem. Now,
let us prove the power allocation problem $\bar{\mathcal{P}}$ is
convex.

It is obvious that $f\left(\boldsymbol{p}\right)$ is a concave function
of $\boldsymbol{p}$. Therefore, we only need to prove $d\left(\boldsymbol{p}\right)\triangleq\textrm{\textrm{ACRB}}\left(\boldsymbol{p}\right)$
is a convex function of $\boldsymbol{p}$. First, the function $d\left(\boldsymbol{p}\right)=\textrm{Tr}\left\{ \left[\mathbb{E}_{\mathbf{\Theta}^{r}}\left[\hat{\mathbf{J}}\left(\boldsymbol{x}\right)\right]+\mathbf{J}^{p}\left(\boldsymbol{\theta}^{r}\right)\right]^{-1}\right\} $
is a convex function of $\mathbb{E}_{\mathbf{\Theta}^{r}}\left[\hat{\mathbf{J}}\left(\boldsymbol{x}\right)\right]$.
Second, every element in $\mathbb{E}_{\mathbf{\Theta}^{r}}\left[\hat{\mathbf{J}}\left(\boldsymbol{x}\right)\right]$
is a convex function of $\boldsymbol{p}$. Therefore, the function
$d\left(\boldsymbol{p}\right)\triangleq\textrm{\textrm{ACRB}}\left(\boldsymbol{p}\right)$
is a convex function of $\boldsymbol{p}$.
\end{IEEEproof}
By Theorem \ref{thm:asy-optimal-distribution}, the Pareto optimal
boundary of the capacity-BCRB region for large $N$ (or equivalently,
the capacity-ACRB region) can be obtained by solving the convex power
allocation problem $\bar{\mathcal{P}}$ with different distortion
constraint $D$. Since the communication-optimal power allocation
is known to have a water-filling structure, in the following section,
we shall focus on study the structure of the S-optimal power allocation
that minimizes the $\textrm{\textrm{ACRB}}\left(\boldsymbol{p}\right)$.

\section{Sensing-optimal Power Allocation}

The S-optimal power allocation can be obtained through the following
convex optimization problem:

\begin{subequations}
\begin{eqnarray}
\mathcal{P}_{s}: & \underset{\boldsymbol{p}}{\min} & d\left(\boldsymbol{p}\right),\\
 & \text{s.t.} & \mathbf{1}^{T}\boldsymbol{p}=P,\\
 &  & 0\leq P_{nm}\leq P_{max},
\end{eqnarray}
\end{subequations}

Note that $\mathcal{P}_{s}$ is still hard to analyze under arbitrarily
prior information $p_{\mathbf{\Theta}^{r}}\left(\boldsymbol{\theta}^{r}\right)$,
so we add some reasonable technical assumptions to derive the following
corollary.
\begin{cor}
\label{cor:Asump_pri}Under the assumption that $\mathbb{E}\left[\alpha_{k}^{r}\right]=0,\forall k$,
$\hat{\mathbf{J}}_{kk}\left(\boldsymbol{x}\right)\triangleq\mathbb{E}_{\mathbf{\Theta}^{r}}\left[\mathbf{L}_{k}^{-T}\tilde{\mathbf{J}}_{kk}^{o}\left(\boldsymbol{x}\right)\mathbf{L}_{k}^{-1}\right],\forall k$
is a diagonal matrix given by (\ref{eq:FIM_avg}), where $\tilde{V_{k}}\triangleq\underset{m}{\sum}\underset{n}{\sum}\left(n-1\right)^{2}P_{nm}$
and $\sigma_{\alpha_{k}^{r}}^{2}\triangleq\mathbb{E}\left[\left|\alpha_{k}^{r}\right|^{2}\right]$.
Moreover, the prior information matrix $\mathbf{J}^{p}\triangleq\textrm{blkdiag}\left[\mathbf{J}_{11}^{p},\cdots,\mathbf{J}_{KK}^{p}\right]$
is a block-diagonal matrix, and $\mathbf{J}_{kk}^{p},\forall k$ are
diagonal matrices determined by the mean vector and covariance matrix
of $\boldsymbol{\theta}^{r}$.
\end{cor}
\begin{IEEEproof}
The proof follows from straightforward calculations and is omitted
due to space limit.
\end{IEEEproof}
Therefore, when $\mathbb{E}\left[\alpha_{k}^{r}\right]=0,\forall k$,
the ACRB is given by (\ref{eq:ACRB_p}), where $\textrm{ACRB}^{\alpha_{k}^{r}}\left(\boldsymbol{p}\right)$
and $\textrm{ACRB}^{\tau_{k}^{r}}\left(\boldsymbol{p}\right)$ are
given by equation (\ref{eq:ACRB_p_alpha}) and (\ref{eq:ACRB_p_tau}),
respectively.

\begin{equation}
\hat{\mathbf{J}}_{kk}\left(\boldsymbol{x}\right)=2\left(\sigma^{r}\right)^{-2}\textrm{diag}\left[P,P,\left(\omega_{0}\sigma_{\alpha_{k}^{r}}\right)^{2}\tilde{V_{k}}\right],\label{eq:FIM_avg}
\end{equation}

\begin{equation}
\textrm{ACRB}\left(\boldsymbol{p}\right)=\underset{k}{\sum}\left[\textrm{ACRB}^{\alpha_{k}^{r}}\left(\boldsymbol{p}\right)+\textrm{ACRB}^{\tau_{k}^{r}}\left(\boldsymbol{p}\right)\right],\label{eq:ACRB_p}
\end{equation}

\begin{equation}
\textrm{ACRB}^{\alpha_{k}^{r}}\left(\boldsymbol{p}\right)\triangleq\frac{1}{\left(\sigma^{r}\right)^{-2}P+\left(\mathbf{J}_{kk}^{p}\right)_{11}},\label{eq:ACRB_p_alpha}
\end{equation}

\begin{equation}
\textrm{ACRB}^{\tau_{k}^{r}}\left(\boldsymbol{p}\right)\triangleq\frac{1}{2\left(\sigma^{r}\right)^{-2}\left(\omega_{0}\sigma_{\alpha_{k}^{r}}\right)^{2}\tilde{V_{k}}+\left(\mathbf{J}_{kk}^{p}\right)_{33}}.\label{eq:ACRB_p_tau}
\end{equation}

\begin{thm}
\label{thm:sensing_opt}The optimal condition for $\mathcal{P}_{s}$
under the assumption $\mathbb{E}\left[\alpha_{k}^{r}\right]=0,\forall k$
is that
\[
P_{nm}=\begin{cases}
P_{max}, & N-N_{act}+1\leq n\leq N,\\
\frac{P_{e}}{M}, & n=N-N_{act},\\
0, & n<N-N_{act},
\end{cases}
\]
where $P_{e}=P-P_{max}MN_{act}$, $N_{act}=\left\lceil \frac{P}{MP_{max}}\right\rceil $
and $\left\lceil \cdot\right\rceil $ is the ceiling operation.
\end{thm}
\begin{IEEEproof}
It can be observed from equation (\ref{eq:ACRB_p}) that the $\textrm{ACRB}\left(\boldsymbol{p}\right)$
is minimized when $2\left(\sigma^{r}\right)^{-2}\left(\omega_{0}\sigma_{\alpha_{k}^{r}}\right)^{2}\tilde{V_{k}}$
is maximized. Theorem \ref{thm:sensing_opt} means that to achieve
the best sensing performance, we must allocate the maximum transmitted
power at the $N_{act}$ edge-most subcarriers .
\end{IEEEproof}

\section{Simulation Results}

In this section, we shall use simulations to compare the communication
and sensing tradeoff by varying $D$ and solving $\bar{\mathcal{P}}$
to obtain the capacity ACRB region $\left(C,D\right)$. In the simulation
figures, we only show the delay estimation error in the x-axis for
clarity. The subcarrier spacing $f_{0}$ is fixed as 15 KHz.

\textcolor{black}{}
\begin{figure}[h]
\centering{}\textcolor{black}{}%
\begin{minipage}[t]{0.45\textwidth}%
\begin{center}
\textcolor{black}{\includegraphics[clip,width=80mm]{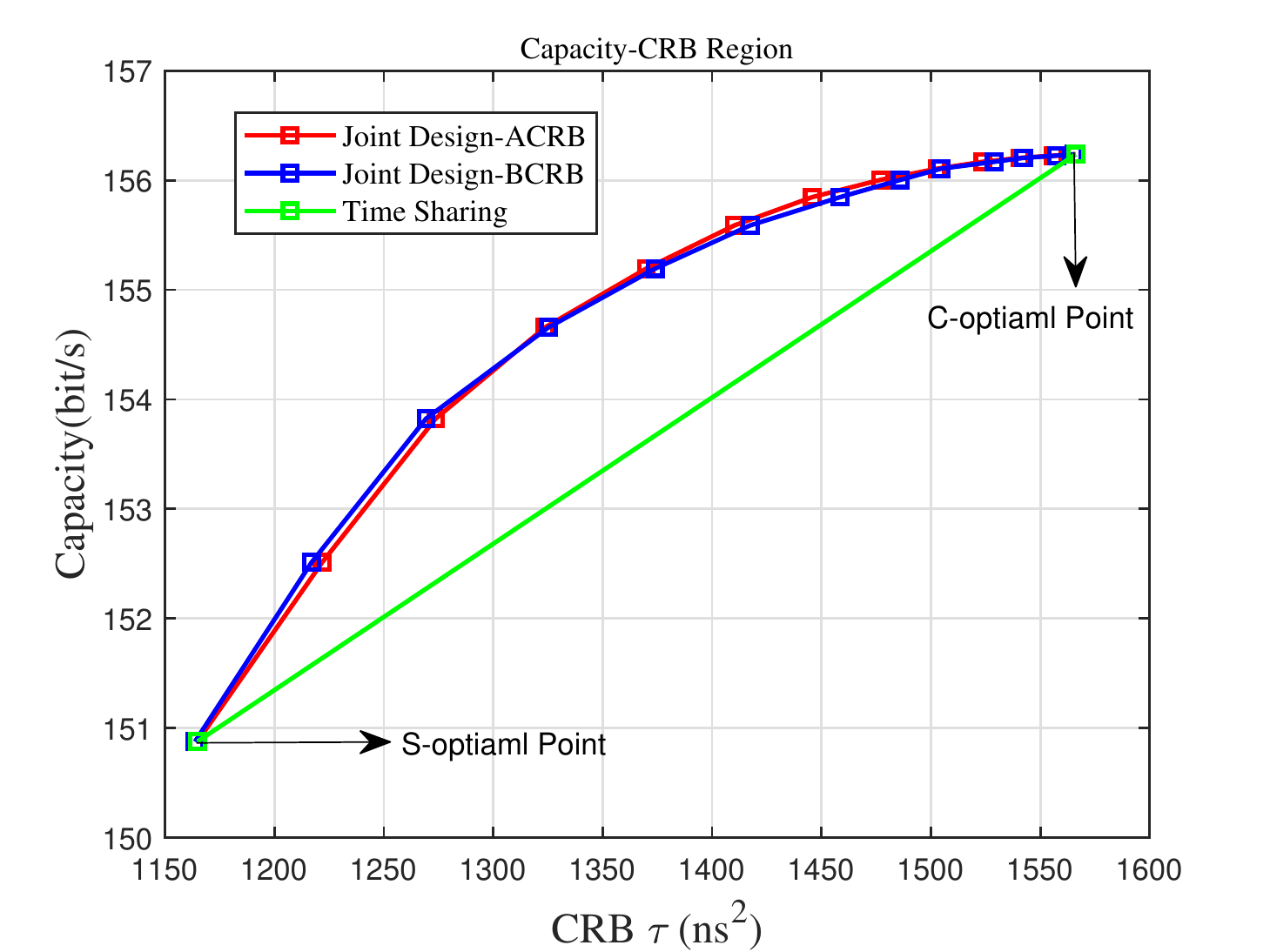}}\vspace{-7mm}
\par\end{center}
\textcolor{black}{\caption{\label{fig:C_D_single}\textcolor{black}{Capacity-ACRB Region for
single target case.}}
}%
\end{minipage}\textcolor{black}{\hfill{}}%
\begin{minipage}[t]{0.45\textwidth}%
\begin{center}
\textcolor{black}{\includegraphics[clip,width=80mm]{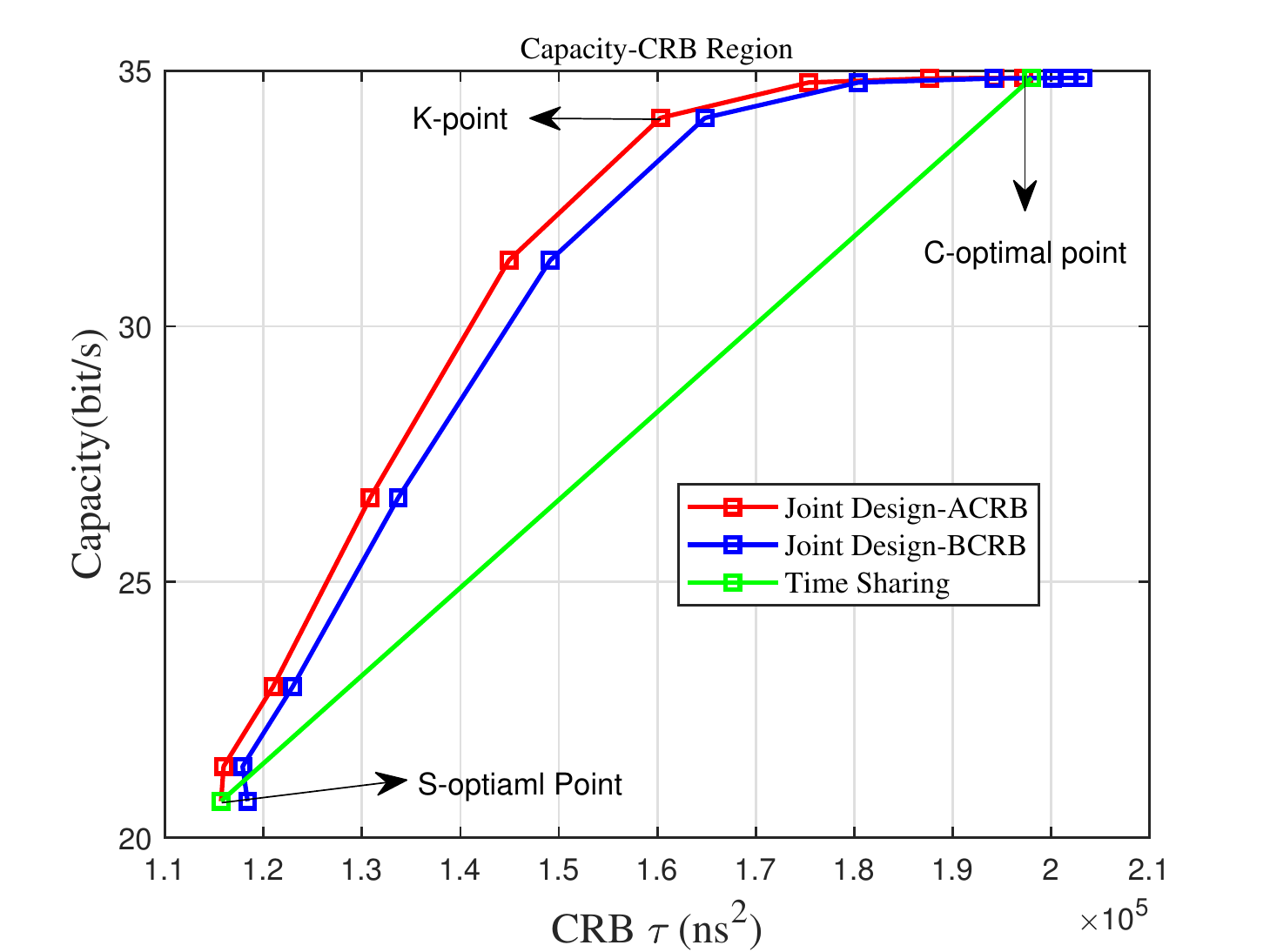}}
\par\end{center}
\vspace{-7mm}

\textcolor{black}{\caption{\label{fig:C_D_multi}\textcolor{black}{Capacity-ACRB Region for two
target case.}}
}%
\end{minipage}
\end{figure}

Fig. \ref{fig:C_D_single} illustrates the Capacity-BCRB\textbackslash ACRB
Region for the single target case. We set $N=1024,M=1$ and the prior
information $p_{\mathbf{\Theta}^{r}}\left(\boldsymbol{\theta}^{r}\right)$
as Gaussian distribution. As can be observed, the joint design scheme
by solving the convex optimization problem $\bar{\mathcal{P}}$ has
significant performance gain compared with the time sharing scheme
which allocating orthogonal resources for sensing and communication
and the ACRB is a good approximation of BCRB for large $N$ with negligible
approximation error. Moreover, when $N$ is large and there is only
a single target, there is almost no tradeoff between communication
and sensing, e.g., the capacity loss of the S-optimal point is only
about 3\% compared to the C-optimal point.

In Fig. \ref{fig:C_D_multi}, we plot the Capacity-BCRB\textbackslash ACRB
Region for the two target case. We set $N=256,M=1$, the prior information
$p_{\mathbf{\Theta}^{r}}\left(\boldsymbol{\theta}^{r}\right)$ as
Gaussian distribution and the difference between the delay of two
targets as $523$ ns, which is $\frac{2}{Nf_{0}}$. As can be observed,
the joint design scheme still has significant performance gain compared
with the time sharing scheme. However, the approximation error of
ACRB is larger than the case when $N=1024$. Moreover, when $N$ is
smaller and there are two close targets, there is a tradeoff between
communication and sensing, e.g., the capacity loss of the S-optimal
point is about 42\% compared to the C-optimal point. In this case,
it is desirable to operate the ISAC system at the knee point where
the communication capacity loss is small and the sensing performance
is also relatively good (e.g., the K-point in Fig. \ref{fig:C_D_multi}).

\section{Conclusions}

In this work, we have investigated the fundamental limit of OFDM ISAC
system. A Capacity-BCRB optimization problem for such system is formulated.
Based on the asymptotic analysis, we show that the asymptotically
optimal input distribution that achieves the Pareto boundary point
of the Capacity-BCRB region is Gaussian and the entire Pareto boundary
can be obtained by solving a convex power allocation problem. Moreover,
we characterize the structure of the sensing-optimal power allocation
in the asymptotically case under some reasonable technical assumptions.
As future work, it is of great interest to extend the current framework
to multi-terminal ISAC topologies (such as MACs and BCs).

\bibliographystyle{IEEEtran}
\bibliography{isac_ofdm}

\end{document}